\documentclass[twocolumn,
english,aps,prl,superscriptaddress,showpacs,preprintnumbers]{revtex4-1}%
\usepackage{graphicx}
\usepackage{verbatim}
\usepackage{natbib}
\usepackage[draft]{hyperref}%
\usepackage{amsmath}%
\usepackage{amsfonts}%
\usepackage{amssymb}
\usepackage{dcolumn}
\usepackage{float}
\providecommand{\U}[1]{\protect\rule{.1in}{.1in}}

\begin{document}
\title{Optomechanical coupling in a two-dimensional photonic crystal defect cavity}
\author{E. Gavartin}
\affiliation{Ecole Polytechnique F{\'e}d{\'e}rale de Lausanne, EPFL, 1015 Lausanne, Switzerland}
\author{R. Braive}
\affiliation{Laboratoire de Photonique et de Nanostructures, Route de Nozay, 91460 Marcoussis, France}
\affiliation{Universit{\'e} Paris Denis Diderot, 75205 Paris, Cedex 13, France }
\author{I. Sagnes}
\affiliation{Laboratoire de Photonique et de Nanostructures, Route de Nozay, 91460 Marcoussis, France}
\author{O. Arcizet}
\affiliation{Institut N{\'e}el, 25 rue des Martyrs, 38042 Grenoble, France}
\author{A. Beveratos}
\affiliation{Laboratoire de Photonique et de Nanostructures, Route de Nozay, 91460 Marcoussis, France}
\author{T. J. Kippenberg}
\email{tobias.kippenberg@epfl.ch}
\affiliation{Ecole Polytechnique F{\'e}d{\'e}rale de Lausanne, EPFL, 1015 Lausanne, Switzerland}
\affiliation{Max Planck Institut f\"ur Quantenoptik, 85748 Garching, Germany}
\author{I. Robert-Philip}
\email{isabelle.robert@lpn.cnrs.fr}
\affiliation{Laboratoire de Photonique et de Nanostructures, Route de Nozay, 91460 Marcoussis, France}

\begin{abstract}
Periodically structured materials can sustain both optical and mechanical
modes. Here we investigate and observe experimentally the optomechanical properties of a
conventional two-dimensional suspended photonic crystal defect cavity with a mode volume of $\sim$$3\left(\lambda/n\right)^{3}$. Two
families of mechanical modes are observed: flexural modes, associated to the motion of the whole suspended membrane, and localized modes with frequencies in the GHz regime corresponding to localized phonons in the optical defect
cavity of diffraction-limited size. We demonstrate direct measurements of the
optomechanical vacuum coupling rate using a frequency calibration technique.
The highest measured values exceed 250 kHz, demonstrating strong coupling of
optical and mechanical modes in such structures.

\end{abstract}

\maketitle







Cavity optomechanics \cite{Kippenberg2008,*Marquardt2009,*Favero2009} exploits the coupling of mechanical oscillators to the light field via
radiation pressure. On the applied side, such coupling may be used to enable novel radiation pressure driven clocks \cite{Kippenberg2005,*Carmon2005}, make
highly sensitive displacement sensors, tunable optical filters \cite{Rosenberg2009, *Wiederhecker2009}, delay lines \cite{Safavi-Naeini2010c} or enable photon storage on a chip \cite{Weis2010, Chang2010}. On a fundamental level,
such systems can be exploited for demonstrating that nano- and micromechanical
oscillators can exhibit quantum mechanical behavior \cite{Schwab2005}. It has been shown experimentally that it is possible to cool a mechanical
oscillator intrinsically via radiation pressure dynamical
backaction \cite{Arcizet2006, *Gigan2006, *Schliesser2006}. In order to reach the ground state of mechanical motion and enable manipulation in the quantum regime, one approach consists
in down-sizing the oscillator, thus shifting the quantum-classical transition
towards higher temperatures. Among various sub-micron
optomechanical systems presently investigated \cite{Thomson2008,*Anetsberger2009,*Li2008},
suspended membranes containing a photonic crystal cavity offer strong light
confinement in diffraction-limited volumes and are therefore natural
candidates for achieving strong optomechanical coupling. Recently,
optomechanical coupling in 1D photonic crystal systems \cite{Foresi1997} has been observed
in patterned single \cite{Eichenfield2009c} and dual
nanobeams (zipper cavities) \cite{Eichenfield2009, Eichenfield2009a}. It would be highly desirable to extend such optomechanical
coupling to 2D systems, notably photonic crystal defect cavities.\ Such
cavities offer the strong light confinement possible (i.e. small mode volume), high quality factor (Q) \cite{Akahane2003} and have been used for studying cavity QED using quantum dots \cite{Yoshie2004} or for realizing nanolasers \cite{Painter1999,Strauf2006}. Recently, a 2D optomechanical photonic
crystal slot cavity has been reported \cite{Safavi-Naeini2010b}. While mechanical displacement due to strong
radiation force generated by band-edge modes in bilayer photonic crystal slabs
has been reported as well \cite{Roh2010}, to date optomechanical
coupling of the 2D conventional defect cavity has not been studied.

In this paper, we demonstrate optomechanical coupling using a photonic crystal
defect (L3) cavity. We provide a direct and robust
experimental determination of the vacuum optomechanical coupling rate \cite{Gorodetsky2010} using frequency
modulation, showing a particularly strong coupling for the localized
mechanical modes, which may also be coupled to quantum dots in future
studies \cite{Wilson-Rae2004}. 

\begin{figure}[h]
\includegraphics[scale=0.29]{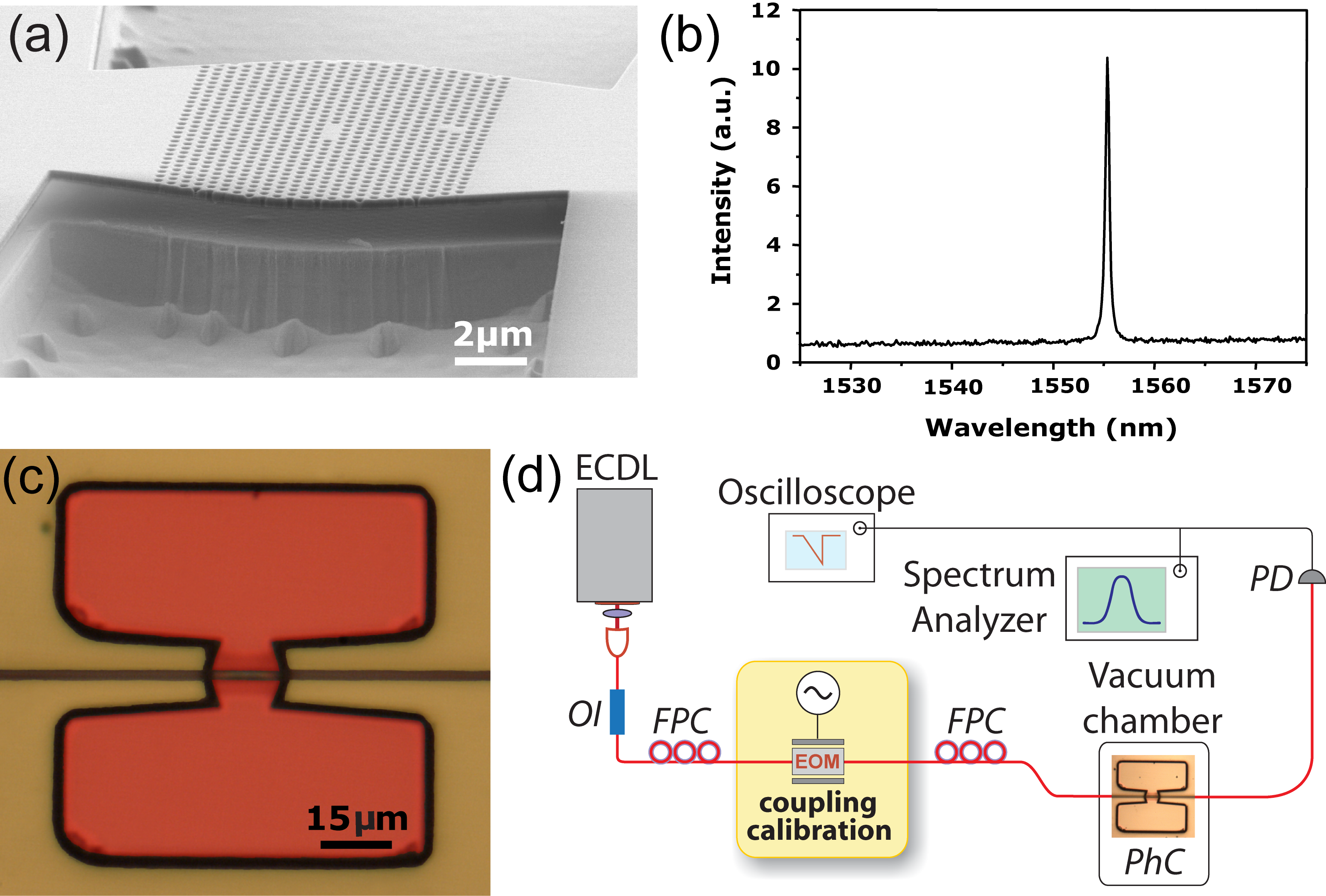} \centering
\caption{(a) Scanning Electron Microscope side view of the cavity. (b)
Microphotoluminescence spectrum of the photonic crystal slab cavity obtained
under non-resonant continuous optical excitation at normal incidence with an excitation power of
100 $\mu$W at 532 nm (c) Micrograph (false colors) of a defect cavity
fiber-taper system used to read out mechanical motion of the cavity. (d)
Experimental setup (ECDL: External cavity diode laser, OI: Optical Isolator, EOM: Electro-optical
modulator, FPC: Fiber polarization controller, PhC: Photonic crystal defect cavity, PD: Photo diode).}%
\label{Fig1}%
\end{figure}

The optomechanical device under study consists of a 262-nm thick InP suspended
membrane containing a two-dimensional photonic crystal defect cavity shown in
Figure \ref{Fig1}(a). The cavity, following the design proposed in
\cite{Akahane2003}, contains three missing holes in a line of a perfect
triangular lattice of holes with a lattice constant of $a=430$ nm and a radius
$r=90$ nm. At both edges of the cavity the holes are displaced outwards by
$d=0.18a$, in order to obtain high optical quality factors. The cavity is
fabricated using electron beam lithography, inductively coupled plasma etching
\cite{Talneau2008}, and wet etching. The cavity incorporates a single layer of
self-assembled InAsP quantum dots \cite{Michon2008} at its vertical center
plane for cavity characterization. The whole structure is grown by metalo-organic chemical vapor
deposition. The quantum dot density is $\sim$$15\times$10$^{9}$
cm$^{-2}$, and their spontaneous emission is centered around 1560 nm at 300 K
with an inhomogeneous broadening of about 150 nm. The presence of the dots
inside the cavity allows to identify the spectral properties of the
fundamental optical mode of the cavity by photoluminescence measurements
\cite{Hostein2010}, as shown in Figure \ref{Fig1}(b). The resonance wavelength
of the fundamental mode is centered around 1555 nm and the cold-cavity quality
factor is measured to be $\sim$$10^{4}$ (cavity linewidth is $\kappa/2\pi\approx 20\;\mathrm{GHz}$). The suspended photonic crystal membrane lies on top of a 10 $\mu$m high mesa
structure (see Fig. \ref{Fig1}(a)). The mesa structure is processed to enable
positioning of a tapered optical fiber in the evanescent field of the cavity,
while precluding any interaction with the nearby substrate. The shape of the
membrane resembles a Bezier curve, which was chosen to increase
optomechanical coupling to the flexural modes. \newline


The setup used in the experiment is depicted in Figure \ref{Fig1}(d). An external-cavity diode laser is used for the readout of the
mechanical motion. Coupling to the optical modes of the suspended
membrane is achieved with the optical fiber-taper technique \cite{Cai2000}.
Piezoelectric actuators enable an accurate positioning allowing to optimize
the gap between the fiber-taper and the defect cavity, and thus to increase
evanescent coupling. Despite careful reduction of the taper-cavity gap, only a
small fraction of the light can be coupled into the defect cavity, typically not exceeding 10 $\%$
of the incoming laser power. The cavity-fiber system is kept in a vacuum
chamber with a pressure below 1 mbar. Laser light coupled inside the photonic crystal cavity leads to local heating, which induces a thermal effect arising from the temperature dependent refractive index $n$ \cite{Carmon2004}. As $dn/dT>0$ ($T$ is the temperature) for InP \cite{Martin1995}, the region detuned to the blue side of the resonance allows thermal passive locking \cite{Carmon2004}. In our experiments the laser frequency is chosen to
correspond to the blue-detuned side of the fringe of the optical mode requiring no further locking.
Mechanical motion of the membrane is imprinted on the transmitted optical
intensity through modulation of the internal cavity field. An electro-optical
modulator is used for frequency modulation to determine the optomechanical
coupling rate. The transmitted signal is detected by a fast receiver and the
electrical signal is analyzed with an oscilloscope as well as an electronic
spectrum analyzer, which is used for the spectral analysis.

\begin{figure}[h]
\includegraphics[scale=0.55]{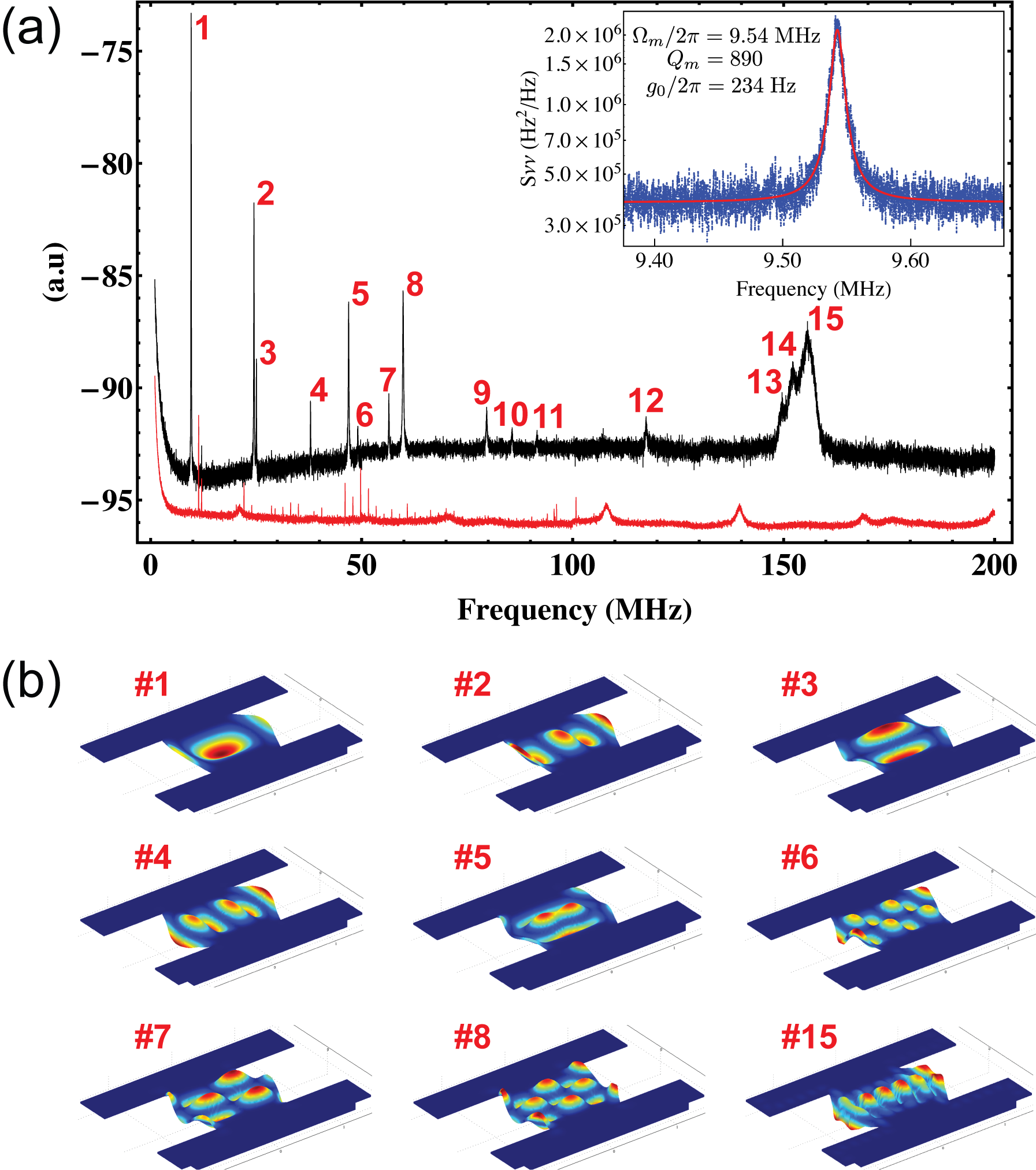} \centering
\caption{(a) Detected frequency noise spectrum in the 1 MHz - 200 MHz range
presenting a series of peaks corresponding to the different mechanical modes
labeled by numbers (Black curve). The red curve represents a spectrum acquired
with the laser being detuned out of resonance. Inset: Calibrated frequency noise spectrum of
the fundamental mode ($\#$1) with a Lorentzian fit (red line). (b) Spatial
displacement pattern of the first eight mechanical modes and the
prominent mode around 150 MHz, as obtained from finite element modeling.}%
\label{Fig2}%
\end{figure}

For a launched laser power of 1.3 mW more than 20 mechanical modes are
observed in the frequency range between 10 MHz and 1 GHz. These modes can be
separated into two mode families. The first family consists of flexural modes
present in the low-frequency range (below 200 MHz), whereas the second family consists of localized modes. Flexural modes, whose spectrum is shown in Figure
\ref{Fig2}(a), correspond to the movement of the whole membrane. In order to
identify the various modes, we modeled the mechanical properties of the
photonic crystal slab structure by finite element modeling (COMSOL
Multiphysics). Realistic geometry parameters were taken into account,
including the under-etching of the mesa structures between which the membrane
is suspended. A good agreement between measurements and modeling is obtained
using a Young modulus of 20 GPa (slightly smaller than usual values observed
in bulk InP materials \cite{ioffe}attributed to the perforation for the photonic crystal). Figure \ref{Fig2}(b) shows the
displacement patterns of the first eight modes as well as a prominent mode
around 150 MHz.

\begin{figure}[h]
\includegraphics[scale=0.7]{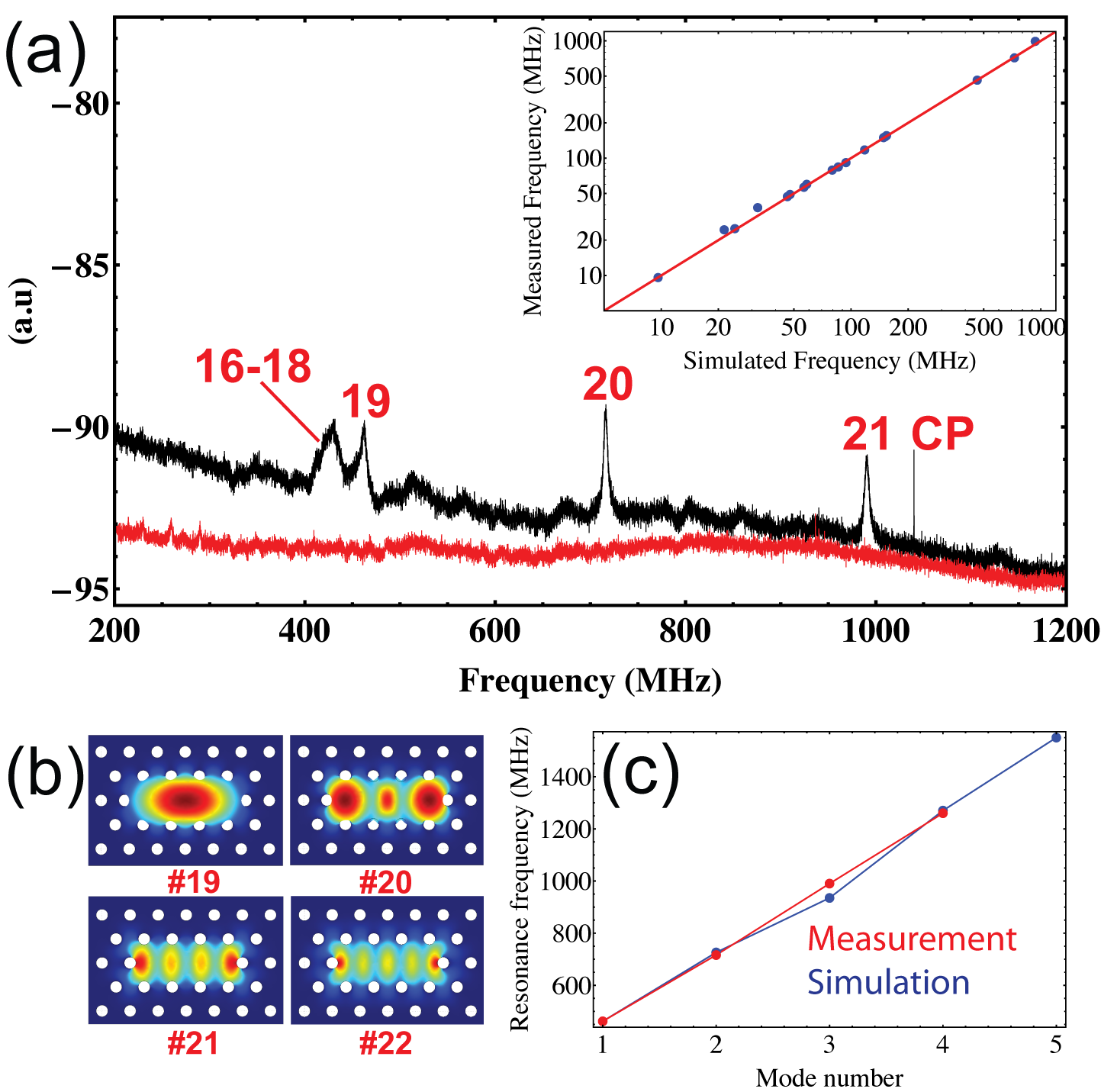} \centering
\caption{(a) Detected frequency noise spectrum in the 200 MHz - 1.1 GHz range
presenting a series of peaks corresponding to the different mechanical modes
labeled by numbers (Black curve). The red curve represents a spectrum acquired
with the laser being detuned out of resonance. CP denotes the calibration peak
resulting from the frequency modulation measurement. Inset: Measured versus simulated frequency for mechanical modes which could be assigned
($\#$1-15 and $\#$19-21). (b) Spatial displacement pattern for
the first four orders of localized mechanical modes, as obtained from finite element
modeling. (c) Simulated (blue) and experimentally determined (red) progression of the resonance frequency of the localized modes versus mode number (mode $\#$22 was determined in a separate measurement).}%
\label{Fig3}%
\end{figure}

Localized modes, shown in Figure \ref{Fig3}, correspond to mechanical
displacement of the membrane localized in the cavity core of the defect. We were able to
resolve the fundamental localized mode at 0.46 GHz as well as the three higher mode orders (at 0.72 GHz, 0.99 GHz and 1.26 GHz). The inset of
Figure \ref{Fig3}(a) compares the simulated against the measured resonance
frequencies of the mechanical modes that could be assigned, revealing excellent agreement both for
flexural and localized modes. The progression of the resonance frequency of the localized modes versus mode number is shown in Figure \ref{Fig3}(c) and follows a linear behavior with mode number.

\begin{figure}[ptb]
\includegraphics[scale=0.5]{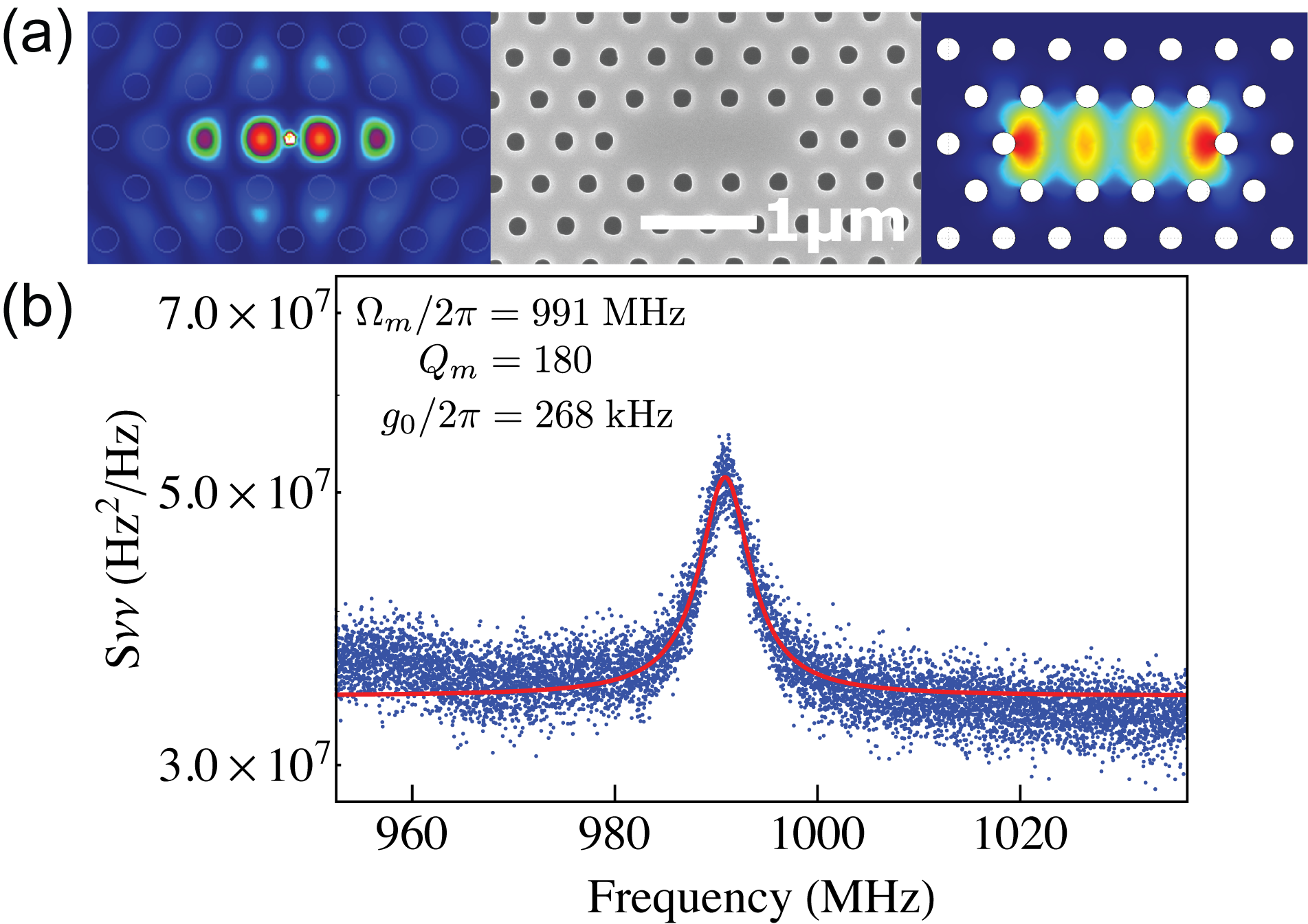} \centering
\caption{(a) Simulated distribution of the electromagnetic field (left),
Scanning Electron Microscope image of the defect (middle) and simulated third
order localized mechanical mode (right). (b) Calibrated frequency noise spectrum of the third
order localized mechanical mode $\#21$ (blue points) with a Lorentzian fit (red
line). A vacuum optomechanical rate of $g_{0}/2\pi=268$ kHz is determined.}%
\label{Fig4}%
\end{figure}

Importantly, localized mechanical modes coincide spatially fully with the optical defect cavity
mode. Therefore, the photonic crystal not only offers strong optical confinement, but simultaneously ultra-high phonon confinement. We note that the localization occurs in the absence of a phononic band gap \cite{Maldovan2006}. Due to the co-localization of the optical and mechanical mode within the defect cavity, strong optomechanical coupling is expected. This is visualized in Figure \ref{Fig4}(a), where the results of the
mechanical FEM simulation for the third order localized mode are compared with
the spatial distribution of the electromagnetic field of the optical defect
mode, which was obtained through a finite difference time-domain (FDTD)
simulation. The high overlap of the mechanical displacement with the
distribution of electromagnetic energy promises strong optomechanical
coupling. In a second experiment we determined the coupling rate for various
flexural modes and the second and third order localized modes.

Usually, the optomechanical coupling strength is determined by two parameters
- the optomechanical coupling parameter $G=\frac{d\omega_{c}}{dx}$, with
$\omega_{c}$ being the resonance frequency of the optical resonator and $x$
denoting the displacement of the mechanical oscillator, and the effective mass
$m_{\mathrm{eff}}$ \cite{Gillespie1995, *Pinard1999} of the mechanical mode.
The necessity of introducing the effective mass routinely arises from the
arbitrary definition of $x$, which often cannot be consistent with the
displacement pattern of different mechanical modes of the system. Particularly
for photonic crystals it is difficult to define an unambiguous displacement
direction of a mechanical mode. One attempt to circumvent this problem is to
introduce an effective length $L_{\mathrm{eff}}$ determined by a perturbative
expression of the overlap of mechanical displacement and the electromagnetic
field distribution \cite{Eichenfield2009, Eichenfield2009a}. Recently, it was
suggested that the vacuum optomechanical coupling rate $g_{0}$ would be a more
proper quantity for optomechanical systems \cite{Gorodetsky2010,
Safavi-Naeini2010}. In analogy to cavity Quantum Electrodynamics (cQED), $g_{0}$ is defined
as $g_{0}=G\cdot x_{\mathrm{zpf}}$, with $x_{\mathrm{zpf}}=\sqrt
{\hbar/2m_{\mathrm{eff}}\Omega_{m}}$ being the zero-point-fluctuations of the
mechanical oscillator ($\hbar$ indicates the reduced Planck constant and
$\Omega_{m}/2\pi$ the mechanical resonance frequency). As all relevant
optomechanical parameters can be derived through knowledge of $g_{0}$,
acquiring its value would make the determination of $G$ and $m_{\mathrm{eff}}$
redundant. As described in \cite{Gorodetsky2010}, the value $g_{0}$ can be
determined experimentally via%

\begin{align}
g_{0}=\sqrt{S_{\omega\omega}\left( \Omega_{m}\right) \frac{\Gamma_{m}}%
{4\bar{n}}},\label{Eq:g0}%
\end{align}

with $\bar{n}$ being the average phonon occupancy of the mechanical mode,
$\Gamma_{m}/2\pi$ being the mechanical damping rate and $S_{\omega\omega}\left(
\Omega_{m}\right) $ being the frequency spectral density of cavity frequency noise evaluated at the
mechanical resonance frequency. For high phonon occupancy one can approximate
$\bar{n}\approx k_{\mathrm{B}}T/\hbar\Omega_{m}\gg1$, with $k_{\mathrm{B}}$ as
Boltzmann's constant. $S_{\omega\omega}\left(
\Omega_{m}\right) $ can be experimentally determined by a frequency modulation
technique \cite{Schliesser2008}.

We performed a measurement of the cavity frequency noise by applying a known phase modulation to the
laser using an LiNbO$_{3}$ electro-optical modulator (cf. Figure \ref{Fig1}). The incoming laser power
was reduced to 0.9 mW. To make sure that the mechanical modes were
not driven thermally, the laser frequency was slightly detuned from the side
of the fringe in both directions just before the experiment. With no change in
the resonance frequency of the mechanical mode occurring, we inferred that the
mechanical modes were only driven by thermal Brownian motion ($T$=300 K). The modulation
frequency was chosen to be close to the resonance frequency of the mechanical
mode to be calibrated. The measurement also allows calibration of the frequency noise produced by the mechanical mode. A detailed
account of the calibration method is given in \cite{SI}. A calibrated frequency noise spectrum
for the fundamental flexural mode is shown in the inset of Figure \ref{Fig2}(a), and
the optomechanical vacuum coupling rate was determined to be $g_{0}/2\pi=
234\;\mathrm{Hz}$. The mechanical quality factor for this mode is $Q_{m}=890$ being the highest for all flexural modes. The coupling rate for the flexural modes increases
with the mode number up to several kHz. The optomechanical coupling for the
localized mechanical modes was determined to be $g_{0}/2\pi=199\;\mathrm{kHz}$ ($Q_{m}=160$)
and $g_{0}/2\pi=268\;\mathrm{kHz}$ ($Q_{m}=180$) for the second and third order,
respectively. A calibrated frequency noise spectrum for third order localized mode is shown in Figure \ref{Fig4}(b). The high values of $g_{0}/2\pi$ give a definite experimental proof of the high optomechanical
coupling between a photonic crystal defect cavity and a localized mechanical
mode. These coupling values are two orders of magnitudes higher than measured in
whispering gallery mode toroidal resonators \cite{Schliesser2008b} and
doubly-clamped strained silicon nitride beams in the near-field of a silica
toroidal resonator \cite{Anetsberger2009}, both of which are ca.
$g_{0}/2\pi\approx1\;\mathrm{kHz}$. Moreover, the measured values are as high as the recently reported coupling of a flexural mode to a photonic crystal slot cavity
\cite{Safavi-Naeini2010c}. The values of $g_{0}/2\pi$ which could be
unambiguously obtained are specified in Table \ref{Table1}. 

The calibration technique used has the clear advantage that it allows to determine $g_{0}$ for any mechanical mode from the mere knowledge of its mechanical linewidth and the mechanical mode occupancy. The method does not require knowledge about the optical linewidth, the specific transduction mechanism of the signal through the optical cavity, the spatial distribution of electromagnetic energy or the displacement pattern of the mechanical mode. Thus the method is particularly suitable for optomechanical calibration in photonic crystals due to the frequently complex spatial distribution of both mechanical and optical modes. Moreover, the present planar architecture can be used for photon-phonon conversion experiments \cite{Safavi-Naeini2010c}.\newline

\begin{table}[ptb]
\caption{Optomechanical vacuum coupling rates for various modes
determined experimentally via the frequency modulation technique.}%
\label{Table1}%
\begin{ruledtabular}
\begin{tabular}{|c|c|c|}
Mode & Measured frequency          &  Measured vacuum  \\
index & $\Omega_{m}/2\pi$ (MHz)&coupling rate $g_{0}/2\pi$ (kHz)\\
\hline
1 & 9.54 & 0.23\\
2 & 24.45 & 0.67\\
5 & 46.87 & 2.26\\
8 & 59.83 & 7.26\\
20 & 716 &  199\\
21 & 991 & 268\\
\end{tabular}
\end{ruledtabular}
\end{table}

In conclusion, we demonstrated optomechanical coupling in a
two-dimensional III-V photonic crystal defect cavity. We observed both
flexural as well as localized mechanical modes. Furthermore, we provide direct
measurements of the vacuum optomechanical coupling rate in a photonic crystal and measure for the first time the coupling of a localized photonic crystal defect mode with a localized 2D mechanical mode. Coupling rates for the localized modes
exceed 250 kHz and are two orders of magnitude larger than in conventional
optomechanical systems. By integrating a single quantum dot in the defect
cavity, a variety of experiments can be envisioned such as laser cooling
\cite{Wilson-Rae2004} and coupling of a quantum mechanical oscillator to an
artificial atom, once the ground state of the mechanical oscillator is reached.

Funding for this work was provided by European NanoSci-ERA project NanoEPR, 
through QNEMS and MINOS by the FP7, the NCCR Quantum Photonics, the SNF and through an ERC Starting Grant SiMP.\newline

%


\newpage
\widetext
\newpage

\renewcommand{\thefigure}{\textbf{S}\arabic{figure}}
\renewcommand{\thetable}{\textbf{S}\arabic{table}}
\renewcommand{\theequation}{$\mathrm{S\,} $\arabic{equation}}
\setcounter {figure} {0}
\setcounter {equation} {0}
\setcounter {table} {0}
\begin{center}
\large{\textbf{
Supplementary information - Optomechanical coupling in a two-dimensional photonic crystal defect cavity}}
\end{center}
\vspace{.2in}

\setcounter{page}{6}

\section{Determination of the optomechanical vacuum coupling rate}

The optomechanical vacuum coupling rate $g_{0}/2\pi$ is determined via a frequency modulation technique. An electro-optical modulator (EOM) is used to phase modulate the laser carrier before its coupling into the cavity. The modulated phase $\Phi$ of the signal can be written as

\begin{eqnarray}
\Phi (t)=\omega_{0} t + \beta \cos (\Omega_{mod} t),
\label{eq:phase}
\end{eqnarray} 
 
with $\omega_{0}$ as the radial laser frequency, $t$ as the time, $\beta$ as the phase shift factor and $\Omega_{mod}/2\pi$ as the modulation frequency. The transduced signal from the photonic crystal cavity is detected with a fast receiver and the spectrum is resolved with an electrical spectrum analyzer (ESA). The spectrum exhibits the Lorentzian spectra of the mechanical modes and a calibration peak at $\Omega_{mod}/2\pi$ resulting from the modulation \cite{Schliesser2008SI,Schliesser2008bSI}. If the modulation frequency is chosen to be close (a couple of mechanical linewidths) to the resonance frequency $\Omega_{m}/2\pi$ of the mechanical mode to be calibrated, $g_{0}$ can be determined as \cite{Gorodetsky2010SI}

\begin{eqnarray}
g_{0}^{2}\approx\frac{1}{2\bar{n}}\frac{\beta^{2}\Omega_{mod}^{2}}{2}\frac{\Gamma_{m}}{4\cdot\mathrm{RBW}}\frac{S_{\omega\omega}\left(\Omega_{m}\right)}{S_{\omega\omega}\left(\Omega_{mod}\right)},
\label{eq:g0totalAS}
\end{eqnarray}

with $\bar{n}$ as the average phonon occupancy, $\Gamma_{m}/2\pi$ as the dissipation rate of the mechanical oscillator, RBW as the resolution bandwidth of the ESA, $S_{\omega\omega}\left(\Omega_{m}\right)$ as the double-sided frequency spectral noise density evaluated at the mechanical resonance frequency and $S_{\omega\omega}\left(\Omega_{mod}\right)$ as the double-sided frequency noise spectral density evaluated at the modulation frequency. Only the proportion between $S_{\omega\omega}\left(\Omega_{m}\right)$ and $S_{\omega\omega}\left(\Omega_{mod}\right)$ is required to be known, and it can be readily obtained from the measured peak power spectral density values of the mechanical and calibration signals. The power spectral density is usually given in dBm in the spectrum of the ESA. The proportion can be found with the following relationship

\begin{eqnarray}
S_{\omega\omega}\left(\Omega_{m}\right)=S_{\omega\omega}\left(\Omega_{mod}\right)\cdot 10^{-\left(P_{mod}-P_{m}\right)/10},
\label{eq:transform}
\end{eqnarray}

with $P_{mod}$ being the power at the peak of the calibration signal expressed in dBm and $P_{m}$ being the power at the peak of the mechanical mode, which is expressed in dBm as well. \\

The frequency modulation technique can be also used to calibrate the spectrum in absolute units. This is done by directly determining $S_{\omega\omega}\left(\Omega_{mod}\right)$ via \cite{Gorodetsky2010SI}

\begin{eqnarray}
S_{\omega\omega}\left(\Omega_{mod}\right)=\frac{\left(\beta\Omega_{mod}\right)^{2}}{4\cdot\mathrm{RBW}}.
\label{eq:Somegaomegamod}
\end{eqnarray}

By using an equivalent scheme depicted in Equation \ref{eq:transform}, the spectrum can be calibrated. The calibration holds true as long as the frequencies of the values to be calibrated are close enough to the resonance frequency of the calibration peak. Otherwise, a generalized transduction coefficient needs to be included \cite{Gorodetsky2010SI}.

\section{Calculation of correction factors for the ESA spectrum}

When measuring electrical signals with an ESA, one usually expresses the power detected at a certain frequency in the logarithmic units of dBm. By averaging one obtains an average of the logarithmic power, which is not equal to the logarithmic expression of the averaged power - the value that is of significance in the measurements. The discrepancy is different for random signal, such as background noise and the mechanical spectrum (resulting from thermal motion), and for coherent signals, such as a calibration peak with a large enough signal-to-noise ratio. For random signals it is known \cite{Agilent} that the measured value is about 2.5 dB below the actual one, whereas for truly coherent sources the measured signal corresponds to the real one. For signals that have contributions both from a coherent source and random signals, such as a calibration peak with a rather low signal-to-noise ratio, the correction factor depends on the magnitude of both contributions. 

Modern ESAs can often compensate this discrepancy internally, however as this function was not applied during our measurements, we need to use the correction described above. In the following we will derive the general expression for the correction of the measured signal.\\

A signal arriving on the ESA can be decomposed in its in-phase ($I$) and out-of-phase ($Q$) part. Spectrum analyzers respond to the magnitude of the signal within their RBW passband \cite{Agilent}. The magnitude of a signal (voltage) $v$ represented by $I$ and $Q$ is given by

\begin{eqnarray}
v=\sqrt{I^{2}+Q^{2}}.
\label{eq:defv}
\end{eqnarray}

The average power arriving on the detector is given by 

\begin{eqnarray}
\bar{P}=\left\langle\frac{v^{2}}{50}\right\rangle=\left\langle\frac{I^{2}+Q^{2}}{50}\right\rangle,
\label{eq:avgp}
\end{eqnarray}

where an input impedance of 50 $\Omega$ is considered (equations are generally dimensionless in this section) and the brackets $\langle\rangle$ imply that the mean value is taken of the expression inside them. In the average mode the ESA averages the logarithmic input signal, thus leading to the following expression for the measured power

\begin{eqnarray}
\bar{P}_{meas}=\left\langle 10\cdot\log\left(\frac{I^{2}+Q^{2}}{50}\right)\right\rangle.
\label{eq:avgpmeas}
\end{eqnarray}

The logarithmic value of the actual power of the signal is given by

\begin{eqnarray}
\bar{P}_{log}=10\cdot\log\left(\bar{P}\right)
\label{eq:avgplog}
\end{eqnarray}

It should be intuitive that the values for $\bar{P}_{log}$ and $\bar{P}_{meas}$ are not equal, just as the log of the average is not the same as the average of the log. 

\vspace{0.2in}

The mean values given in Equations \ref{eq:avgp} and \ref{eq:avgpmeas} can be determined by multiplying the variable being measured with its probability density function (PDF) and integrating over the possible values of the variable. The PDFs of the quadratures $I$ and $Q$ can be assumed to be Gaussian with a certain variance $\sigma_{I}$ for $I$ and $\sigma_{Q}$ for $Q$, and a certain offset of the distribution from 0 being $\mu_{I}$ for $I$ and $\mu_{Q}$ for $Q$. In general, one can assume the variance to be the same for both quadratures, so that $\sigma_{I}=\sigma_{Q}=\sigma$. For random signals it holds that $\mu_{I}=\mu_{Q}=0$, whereas for a coherent signal one has $\mu_{Q}=0$, but $\mu_{I}=\mu > 0$. 

\vspace{0.2in}

Using the assumptions stated above, one can write

\begin{eqnarray}
\mathrm{PDF}\left(I\right)=\frac{1}{\sqrt{2\pi}\sigma}\exp\left(-\frac{\left(I-\mu\right)^{2}}{2\sigma^{2}}\right)
\label{eq:pdfi}
\end{eqnarray}

and

\begin{eqnarray}
\mathrm{PDF}\left(Q\right)=\frac{1}{\sqrt{2\pi}\sigma}\exp\left(-\frac{Q^{2}}{2\sigma^{2}}\right).
\label{eq:pdfq}
\end{eqnarray}

From the Equations \ref{eq:avgp} and \ref{eq:avgpmeas} it is obvious that we are interested in the mean values of $v^{2}$. This requires one to find the PDFs for this variable from the PDFs of $I$ and $Q$. This is accomplished by using a general relationship, which links the PDFs $f_{X_{i}}\left(x_{i}\right)$ of independent random variables $X_{i},i=1,2,...n$ to the PDF $f_{Y}\left(y\right)$ of some variable $Y=G\left(X_{1},X_{2},...X_{n}\right)$. This relationship is given by

\begin{eqnarray}
f_{Y}\left(y\right)=\int_{-\infty}^{\infty}\int_{-\infty}^{\infty}...\int_{-\infty}^{\infty}d x_{1}d x_{2}...d x_{n}f_{X_{1}}\left(x_{1}\right) f_{X_{2}}\left(x_{2}\right) ... f_{X_{n}}\left(x_{n}\right)\delta\left(t-G\left(x_{1},x_{2},...x_{n}\right)\right),
\label{eq:genpdf}
\end{eqnarray}

where $\delta$ denotes Dirac's delta function. 

\vspace{0.2in}

Using Equation \ref{eq:genpdf} one can determine the PDF of $v^{2}$ to be

\begin{eqnarray}
\mathrm{PDF}\left(v^{2}\right)=\frac{1}{2\sigma^{2}}\exp\left(-\frac{v^{2}+\mu^{2}}{2\sigma^{2}}\right)\cdot J_{0}\left(\frac{v\mu}{\sigma^{2}}\right),
\label{eq:pdfv2}
\end{eqnarray}

with $J_{0}\left(\frac{v^{2}\mu}{\sigma^{2}}\right)$ as a Bessel function of the first kind.

Using this expressions for the PDF of $v^{2}$, one can calculate $\bar{P}$ and $\bar{P}_{meas}$ by using Equations \ref{eq:avgp} and \ref{eq:avgpmeas}. For $\bar{P}$ one obtains

\begin{eqnarray}
\bar{P}\left(\mu,\sigma\right)&=&\left\langle\frac{v^{2}}{50}\right\rangle=\int_{0}^{\infty}dv^{2}\,\frac{v^{2}}{50}\cdot\mathrm{PDF}\left(v^{2}\right)\notag\\
&=&\frac{\mu^{2}+2\sigma^{2}}{50},
\label{eq:afgppdfvp}
\end{eqnarray}

and for $\bar{P}_{meas}$ one obtains 

\begin{eqnarray}
\bar{P}_{meas}\left(\mu,\sigma\right)&=&\left\langle 10\cdot\log\left(\frac{v^{2}}{50}\right)\right\rangle\notag\\
&=&\int_{0}^{\infty}dv\,10\cdot\log\left(\frac{v^{2}}{50}\right)\cdot\mathrm{PDF}\left(v^{2}\right)\notag\\
&=&\frac{10\cdot\left(\Gamma\left(0,\frac{\mu^{2}}{2\sigma^{2}}\right)+\ln\left(\frac{\mu^{2}}{50}\right)\right)}{\ln\left(10\right)},
\label{eq:afgplogpdfvnew}
\end{eqnarray}

where $\Gamma\left(0,\frac{\mu^{2}}{2\sigma^{2}}\right)$ is an incomplete Gamma function and $\ln$ denotes the natural logarithm.

For random signals we set $\mu$ to 0 obtaining

\begin{eqnarray}
\bar{P}_{meas}\left(\mu=0,\sigma\right)&=&\lim_{\mu\to 0}\bar{P}_{meas}\left(\mu,\sigma\right)=\lim_{\mu\to 0}\frac{10\cdot\left(\Gamma\left(0,\frac{\mu^{2}}{2\sigma^{2}}\right)+\ln\left(\frac{\mu^{2}}{50}\right)\right)}{\ln\left(10\right)}\notag\\
&=&-\frac{10\left(\Gamma_{E}+\ln\left(\frac{25}{\sigma^{2}}\right)\right)}{\ln\left(10\right)},
\label{eq:afgpmeasmu0}
\end{eqnarray}

with $\Gamma_{E}$ as the Euler-Mascheroni constant ($\Gamma_{E}\approx 0.577216$). 

Finally, we obtain a general expression for the discrepancy between the actual and the measured signal

\begin{eqnarray}
\Delta\left(\mu,\sigma\right)=\bar{P}_{meas}-10\cdot\log\left(\bar{P}\right)=\frac{10\cdot\left(\Gamma\left(0,\frac{\mu^{2}}{2\sigma^{2}}\right)+\ln\left(\frac{\mu^{2}}{\mu^{2}+2\sigma^{2}}\right)\right)}{\ln\left(10\right)}.
\label{eq:generalerror}
\end{eqnarray}

For the special case of a random signal the final expression reads

\begin{eqnarray}
\Delta\left(\mu=0,\sigma\right)=\lim_{\mu\to 0}\Delta\left(\mu,\sigma\right)=-\frac{10\cdot\Gamma_{E}}{\ln\left(10\right)},
\label{eq:randomerror}
\end{eqnarray}

The discrepancy in this case is a constant value being approximately $\Delta\left(\mu=0,\sigma\right)=-2.50682$ and independent of $\sigma$ as expected. 

For a strongly coherent signal with $\mu\gg\sigma$ one obtains $\Delta\left(\mu\gg\sigma,\sigma\right)=0$. This means that for a strong coherent drive there is no discrepancy between the measured and the real one.

\vspace{0.2in}

\section{Experimental Implementation}

\vspace{0.2in}

\begin{figure}[ptb]
   \centering
   \includegraphics[scale=0.4]{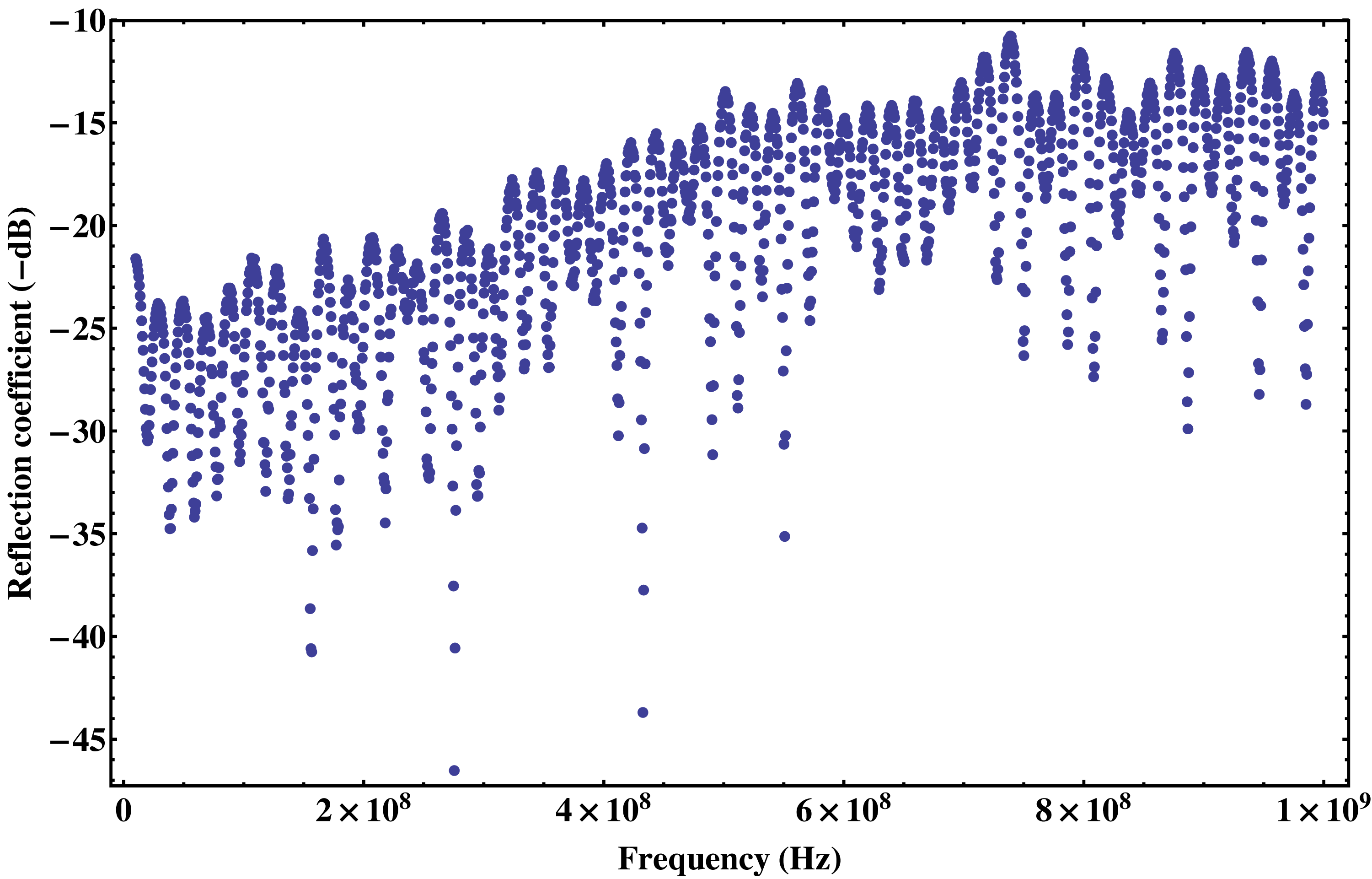}
   \caption{Reflection coefficient $\Gamma_{EOM}$ of the power being received by the combined EOM+cable system versus frequency measured with a network analyzer.}
    \label{EOMRefl}
\end{figure}

\begin{figure}[ptb]
   \centering
   \includegraphics[scale=0.4]{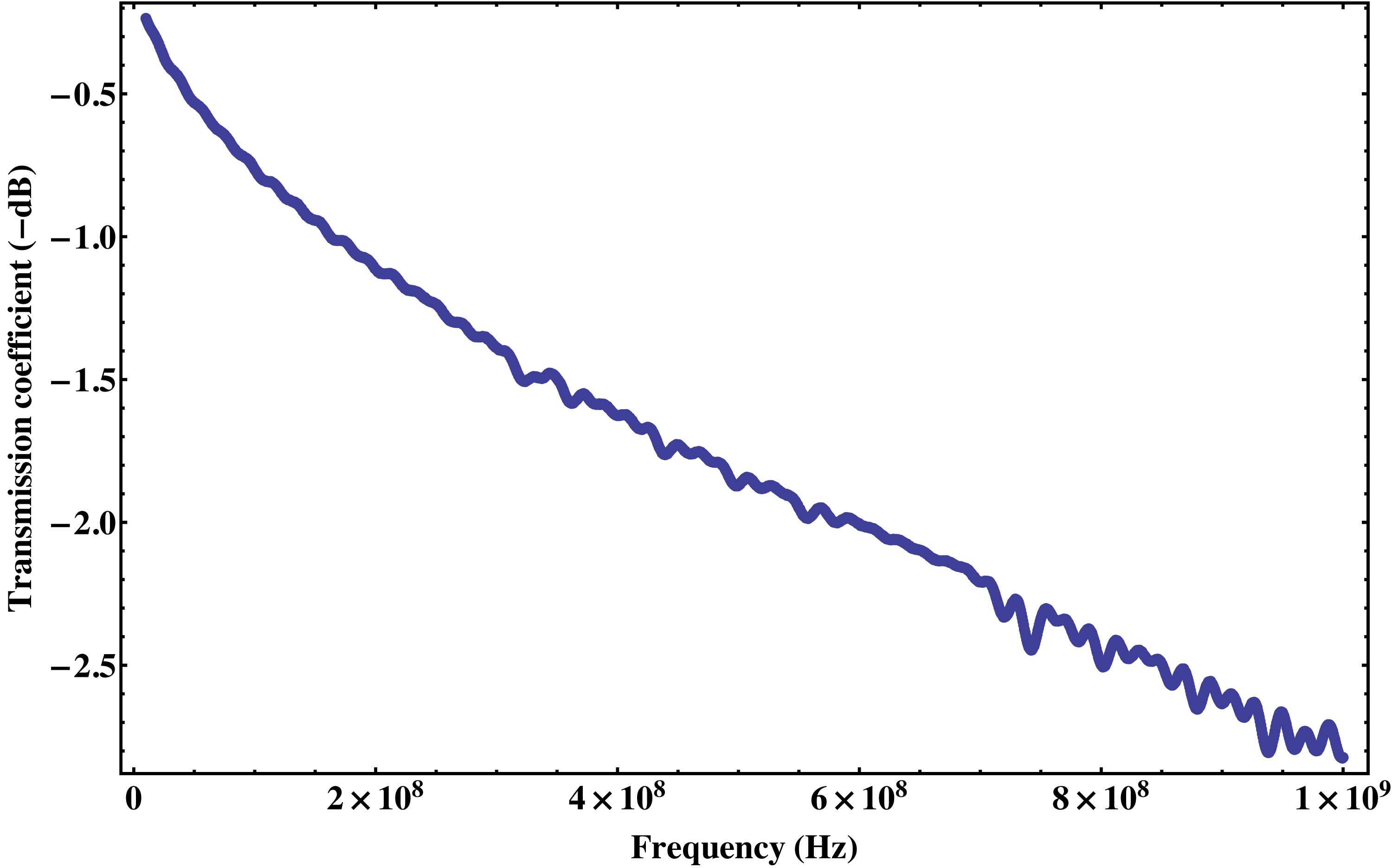}
   \caption{Transmission coefficient $T_{BNC}$ of the power being transmitted through the BNC cable linking the signal generator and the EOM versus frequency. The measurement was performed with a network analyzer.}
    \label{CableTrans}
\end{figure}

The phase shift factor $\beta$ can be determined by 

\begin{eqnarray}
\beta=V/V_{\pi}\cdot \pi=\sqrt{2P_{rf}Z}\pi/V_{\pi},
\label{eq:beta}
\end{eqnarray}

with $V$ as the applied voltage, $V_{\pi}$ as the voltage needed to induce a phase shift of $\pi$, $P_{rf}$ as the input power of the signal generator providing the modulation signal and $Z$ as the impedance between the signal generator and the EOM. To have a quantitavie understanding of the actual voltage applied on the EOM, we directly measured the reflectance of the system consisting of the EOM and the BNC cable linking it to the signal generator. Moreover, we measured transmittance through the BNC cable to account for potential losses. The measurements were performed using a network analyzer.  Figure \ref{EOMRefl} shows the fraction of the reflected power $\Gamma_{EOM}$ of the combined EOM+cable system as a function of the output frequency of the signal generator. Figure \ref{CableTrans} shows the fraction of transmitted power $T_{BNC}$ through cable as a function of frequency. It is obvious that for high frequencies the power loss becomes fairly strong, which is due to radiation losses in the BNC cable. This fact is undermined by Figure \ref{CableTrans} showing the reflection coefficient $\Gamma_{BNC}$ through the cable, which is well below the one of the total EOM+cable system. 

\begin{figure}[ptb]
   \centering
   \includegraphics[scale=0.4]{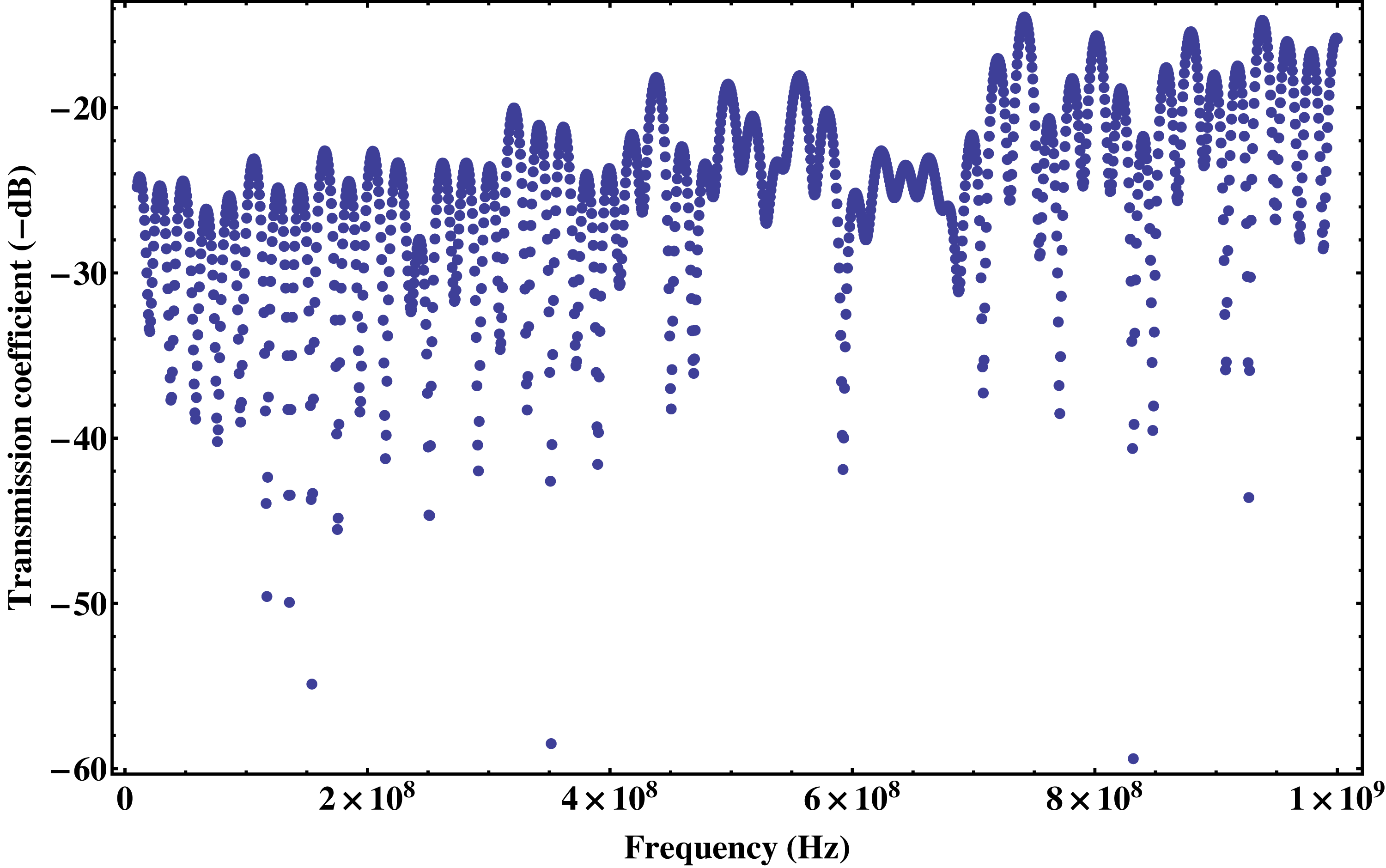}
   \caption{Reflection coefficient $\Gamma_{BNC}$ of the power being transmitted through the BNC cable linking the signal generator and the EOM versus frequency. The measurement was performed with a network analyzer.}
    \label{CableRefl}
\end{figure}

The factor $\beta$ can be thus determined as

\begin{eqnarray}
\beta=\sqrt{2P_{rf}\cdot\left(1-10^{\Gamma_{EOM}/10}\right)\cdot 10^{T_{BNC}/10}\cdot Z_{out}}\pi/V_{\pi}.
\label{eq:measbeta}
\end{eqnarray}

Measurements gave a value of $V_{\pi}\approx7\;\mathrm{V}$ which is consistent with the value provided of the manufacturer of the EOM. 

\vspace{0.2in}

\begin{figure}[ht!]
   \centering
   \includegraphics[scale=0.6]{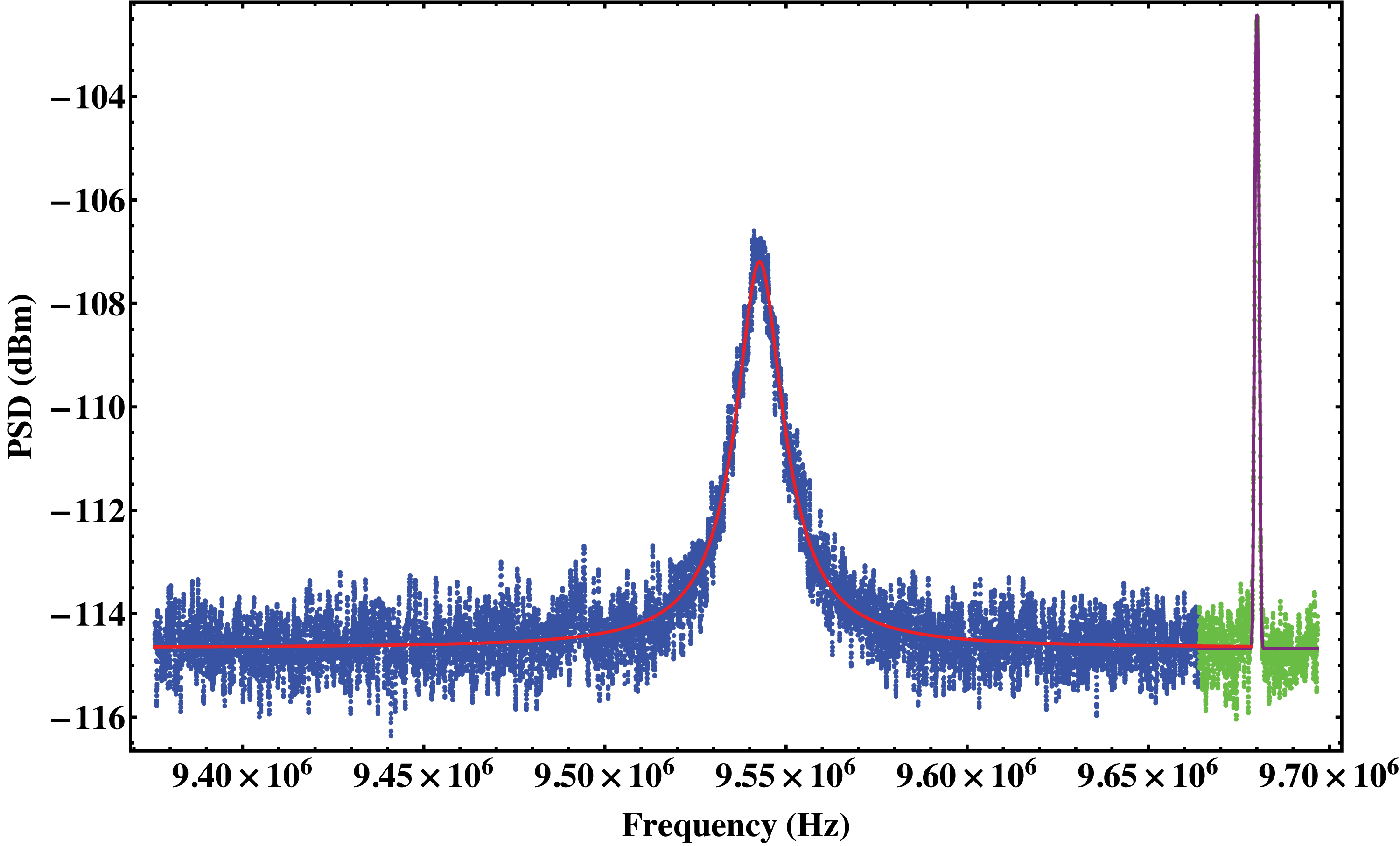}
   \caption{Mechanical mode (blue) with a calibration peak (green) including the respective fits.}
    \label{firstmode}
\end{figure}

A spectrum of the fundamental flexural mode is shown in Figure \ref{firstmode} together with the calibration peak. As discussed in the previous section, the measured spectrum needs to be readjusted. This is done by first fitting the mechanical with a Lorentzian and the calibration peak with a Gaussian. The noise level is obtained from the baseline of the Lorentzian, and this value is subsequently used to determine the variance $\sigma$ as defined in Equations \ref{eq:pdfi} and \ref{eq:pdfq} through use of Equation \ref{eq:afgpmeasmu0}. From the Gaussian fit we obtain the maximum value of the calibration peak. Using this value together with the determined value of $\sigma$, one can numerically solve Equation \ref{eq:afgplogpdfvnew} to obtain $\mu$. The value of $\mu^{2}/50$ is the corrected maximum value of the calibration peak. The spectrum of the random signals, consisting of the background noise and the mechanical response, is corrected with the constant value given in Equation \ref{eq:randomerror}. The corrected values for the maxima of the mechanical spectrum and the calibration peak are used together with $\Gamma_{m}$, inferred from the Lorentzian fit, to determine $g_{0}$ as described in Equation \ref{eq:g0totalAS}. By using Equation \ref{eq:Somegaomegamod} one can calibrate the spectrum in absolute frequency units.

\vspace{0.2in}

The values of $g_{0}/2\pi$, that could be unambiguously determined, are summarized in Table \ref{table1}. The value of $g_{0}/2\pi$ without the correction for logarithmic averaging is also stated together with the calculated correction for the calibration peak. The required correction for the calibration peak is fairly small for most measurements due to the large enough signal-to-noise ratio. In those cases the main correction stemmed from the adjustment of the random signals by a constant factor of around 2.5 dBm.

\begin{table}[H]
\begin{center}
\begin{tabular}{|l|l|l|p{1.5in}|p{1in}|}
  \hline
  Mode  & $\Omega_{m}/2\pi$  & $g_{0}/2\pi$ & $g_{0}/2\pi$ (without correction) & Correction factor for the calibration peak\\
  \hline
  1 & $9.54$ MHz& $234$ Hz& $175$ Hz& 0.000 dBm\\
  2 & $24.48$ MHz& $668$ Hz& $500$ Hz& 0.004 dBm\\
  3 & $46.87$ MHz& $2.26$ kHz& $1.69$ kHz& 0.007 dBm\\
  4 & $59.83$ MHz& $7.26$ kHz& $5.44$ kHz& 0.000 dBm\\
  5 & $716.39$ MHz& $199.1$ kHz& $148.4$ kHz& 0.045 dBm\\
  6 & $990.88$ MHz& $267.8$ kHz& $193.4$ kHz& 0.322 dBm\\ \hline
\end{tabular}
\caption{Table with $g_{0}$ for different mechanical modes as well as different calculation conditions. The last column specifies the correction factor for the peak height of the calibration peak.} 
\label{table1}
\end{center}
\end{table}

The values of $g_{0}/2\pi$, that could be unambiguously determined, are summarized in Table \ref{table1}. The value of $g_{0}/2\pi$ without the correction for logarithmic averaging is also stated together with the calculated correction for the calibration peak. The required correction for the calibration peak is fairly small for most measurements due to the large enough signal-to-noise ratio. In those cases the main correction stemmed from the adjustment of the random signals by a constant factor of around 2.5 dBm.

\end{document}